\documentstyle[preprint,eqsecnum,aps,epsf]{revtex}
\newif\iftightenlines\tightenlinesfalse
\tightenlines\tightenlinestrue
\begin{document}
%
\def\pT{p_T^{\phantom{7}}}
\def\MW{M_W^{\phantom{7}}}
\def\ET{E_T^{\phantom{7}}}
\def\bh{\bar h}
\def\lm{\,{\rm lm}}
\def\lo{\lambda_1}                                              
\def\lt{\lambda_2}
\def\pslt{p\llap/_T}
\def\eslt{E\llap/_T}
\def\to{\rightarrow}
\def\Re{{\cal R \mskip-4mu \lower.1ex \hbox{\it e}}\,}
\def\Im{{\cal I \mskip-5mu \lower.1ex \hbox{\it m}}\,}
\def\SU{SU(2)$\times$U(1)$_Y$}
\def\te{\tilde e}
\def\tl{\tilde l}
\def\ttau{\tilde \tau}
\def\tg{\tilde g}
\def\tga{\tilde \gamma}
\def\tnu{\tilde\nu}
\def\tell{\tilde\ell}
\def\tq{\tilde q}
\def\tw{\widetilde W}
\def\tz{\widetilde Z}
\def\cmsec{{\rm cm^{-2}s^{-1}}}

\hyphenation{mssm}
\def\ds{\displaystyle}
\def\ts{${\strut\atop\strut}$}
%
%
\preprint{\vbox{\baselineskip=14pt%
   \rightline{FSU-HEP-940310}\break 
   \rightline{UH-511-786-94}
}}
\title{TRILEPTONS FROM CHARGINO-NEUTRALINO PRODUCTION\\
AT THE CERN LARGE HADRON COLLIDER}
\author{Howard Baer$^1$, Chih-hao Chen$^1$, Frank Paige$^2$\\
 and Xerxes Tata$^3$}
\address{
$^1$Department of Physics,
Florida State University,
Tallahassee, FL 32306 USA
}
\address{
$^2$Brookhaven National Laboratory, 
Upton, NY 11973 USA
}
\address{
$^3$Department of Physics and Astronomy,
University of Hawaii,
Honolulu, HI 96822 USA
}
\date{\today}
\maketitle
\begin{abstract}
We study direct production of charginos and neutralinos 
at the CERN Large Hadron Collider. 
We simulate all channels of chargino and neutralino production 
using ISAJET 7.07.
The best mode for observing such processes appears to be 
$pp\to\tw_1\tz_2\to 3\ell +\eslt$. We evaluate signal expectations and 
background levels, and suggest cuts to optimize the signal. The trilepton
mode should be viable provided $m_{\tg}\alt 500-600$~GeV; above this mass,
the decay modes $\tz_2\to\tz_1 Z$ and  $\tz_2\to H_{\ell}\tz_1$ 
become dominant, spoiling the signal.
In the first case, the leptonic branching fraction for $Z$
decay is small and additional background from $WZ$ is present, 
while in the second case, the trilepton signal
is essentially absent. For smaller values of $m_{\tg}$, the trilepton signal
should be visible above background, especially if $|\mu|\simeq m_{\tg}$
and $m_{\tell}\ll m_{\tq}$, in which case the leptonic decays of $\tz_2$
are enhanced.
Distributions in dilepton mass
$m(\ell\bar{\ell})$ can yield direct information on neutralino masses
due to the distribution cutoff at $m_{\tz_2}-m_{\tz_1}$. Other distributions
that may lead to an additional constraint amongst the chargino and neutralino
masses are also examined.
\end{abstract}
\medskip
\pacs{PACS numbers: 14.80.Ly, 13.85.Qk, 11.30.Pb}
%
%
%
\section{Introduction}

The search for supersymmetric (SUSY) particles is one of the major issues
in particle physics today\cite{MSSM}. 
Direct searches for SUSY particles at the LEP
$e^+e^-$ collider have led to mass bounds\cite{LEP},
\begin{eqnarray}
m_{\tq}, m_{\tw_1}, m_{\tell} \agt 40\,\hbox{--}\,45\ {\rm GeV},
\eqnum{1}
\end{eqnarray}
where prospects for higher mass searches are linked to increases in
beam energy. At hadron colliders, most searches have focussed on
gluino and squark production; here, the  CDF and D0 experiments have 
obtained mass limits of\cite{CDF,DZERO}
\begin{eqnarray}
m_{\tq}, m_{\tg} \agt 100\,\hbox{--}\,150\ {\rm GeV},
\eqnum{2}
\end{eqnarray}
based on non-observation of events with missing transverse energy
($\eslt$) plus jets above expected background levels. 

Recently, much attention has focussed on the clean trilepton signal
from $p\bar{p}\to\tw_1\tz_2 X$, where $\tw_1\to\ell\nu\tz_1$ and
$\tz_2\to\ell\bar{\ell}\tz_1$. One expects events containing three
hard, isolated leptons plus $\eslt$ with jet activity
only from QCD radiation; standard model (SM)
backgrounds
are expected to be tiny. The importance of this signature has been
pointed out long ago for {\it on shell} $W$ decays\cite{BHT}; it was 
then pointed out that the total $\tw_1\tz_2$ cross section remains 
substantial even for {\it off-shell} $W$ decays\cite{NATH}, so that
the trilepton signal may be observable with an accumulated data
sample of $\sim 100$ $pb^{-1}$. 
Subsequently, it was shown that there could be a large enhancement
of the trilepton signal\cite{BT}, especially when
$m_{\tell} \ll m_{\tq}$ as is the case in the
``no-scale'' limit of supergravity (SUGRA) models. 
In favorable cases, given sufficient luminosity,
it may be possible for Tevatron $p\bar p$ collider experiments to
probe chargino masses even beyond the reach of LEP 200, corresponding to 
gluino masses in the several hundred GeV region. This has since been 
confirmed by
calculations of the
trilepton rate within the
no-scale flipped $SU(5)\times U(1)$ supergravity 
framework\cite{NANOP}. Detailed simulation of the trilepton
signal and background have been carried out in Ref.\cite{BKT}, where the
importance of the dilepton invariant mass distribution was stressed for
obtaining a measurement of $m_{\tz_2}-m_{\tz_1}$. Very recently, 
the CDF\cite{CDFT} and D0\cite{DZERO}
collaborations have reported preliminary bounds on $m_{\tw_1}$ from
a non-observation of trilepton events in their analysis of $\sim 10 pb^{-1}$
of their data. Although these analyses do not (yet) significantly improve
on the bounds from LEP, they clearly establish the viability 
of this signature.

The higher energy ($\sqrt{s}=14$ TeV) and higher luminosity
($10^{33}$--$10^{34}\,\cmsec$) anticipated for the CERN Large
Hadron Collider (LHC) project should substantially increase the range of
parameter space to be probed via the clean trilepton signal. This was
first examined in Ref.\cite{ABNP}, and later in greater detail, by
Barbieri {\it et. al.}\cite{BARB}. These authors\cite{BARB} warned
that at the LHC
the SM background from $t\bar{t}$ production --- largely
negligible at Tevatron energies --- may remain problematic (owing to the large
top pair total cross section), especially if $m_t$ was around 120--130 GeV.
For example, in going from Tevatron to LHC colliders, the $\tw_1\tz_2$ 
production
cross section increases by a factor of $\sim 10$, while total 
$t\bar t$ background cross section ($m_t =175$ GeV) increases
by a factor of $\sim 160$.
In their study, however, they had assumed that the branching
fractions for neutralinos were the same as those of the $Z$ boson
which, as we have mentioned, often leads to an underestimate of
the signal. On the other hand, when probing higher mass scales associated 
with the LHC, possible new chargino and neutralino decay modes may open
up, leading to loss of signal. It is also possible that other
chargino and neutralino reactions,
{\it e.g.} $\tw_1\tz_3$ production, become accessible at the larger LHC energy
and also contribute to the signal.
Finally, if $\tw_1\tz_2$ events can be
isolated from other sources of trileptons,
the high event rate expected may allow for substantial precision in 
the $m_{\tz_2}-m_{\tz_1}$ mass measurement, assuming a signal is found.

In this paper, we seek to answer the following questions:
\begin{enumerate}
\item Can one find a set of cuts to allow a signal to be claimed above
SM backgrounds?
\item If so, in what regions of parameter space is a signal likely observable?
\item Is it possible to separate the trilepton signal from direct
chargino-neutralino production from the same signal coming 
from the cascade decays of gluinos and squarks?
\item Can one gain information on the chargino and neutralino masses?
\end{enumerate}
To answer these, we perform detailed simulations of signal and background
using ISAJET 7.07\cite{ISAJET}. In Sec.~II, we present an overview of total
production cross sections, relevant branching fractions, and details
of our simulation. In Sec.~III, we try to find an optimal set of cuts 
to enhance signal over background, and we outline the regions of parameter
space where a detectable signal can be expected at the LHC. 
We study strategies for 
extracting information about chargino and neutralino masses in Sec.~IV.
We show that for regions
of parameters where $\tz_2 \to \tz_1H_{\ell}$ or $\tz_2 \to \tz_1Z$
decays are kinematically inaccessible, it should be possible to
obtain $m_{\tz_2}-m_{\tz_1}$ with reasonable precision. 
We also discuss the possibility of other mass measurements. We conclude 
in Sec.~V with discussion of our results.

\section{Chargino and neutralino production, decay and event simulation}

We work within the framework of the Minimal Supersymmetric
Model (MSSM)\cite{MSSM}, which is the simplest supersymmetric
extension of the SM. Our MSSM mass and parameter choices are inspired,
but not ruled by, supergravity models with electroweak symmetry breaking: in
particular, we take the higgsino mass parameter $\mu$ as a free parameter,
although it usually scales with $m_{\tg}$, in SUGRA models\cite{AN,SUGRA}
with radiative breaking of electroweak symmetry.
In minimal supergravity models, supersymmetry breaking leads to a common
mass for sfermions at the unification scale. The degeneracy of sfermions 
present at the unification scale
is broken when these masses are evolved down to the weak scale. 
We therefore 
assume slepton masses are related to $m_{\tq},\ m_{\tg}$ and $\tan\beta$
as given by the renormalization group equation (RGE) solutions
listed in Ref. \cite{BT}.
Thus, for $m_{\tilde{q}} \gg m_{\tilde{g}}$,
the squarks are basically degenerate with the sleptons; significant
splitting between the masses of the sleptons and squarks is possible
only when squarks and gluinos are roughly degenerate, in which case
sleptons are considerably lighter than squarks: this latter situation
is frequently realized in ``no-scale'' models\cite{DIMITRI}, in
which neutralino decays to leptons can be enhanced\cite{BT,NANOP}. However,
the trilepton signal may be considerably reduced when decays $\tw_1 \to
\ell\tnu$ or $\tz_2 \to \tell_R\bar{\ell}+\bar{\tell_R}\ell$ 
because the daughter lepton tends to be
soft, reducing the efficiency for passing cuts. 

Pair production of charginos and neutralinos at hadron colliders takes
place via $pp\to\tw_i\tz_jX$ (eight reactions), 
$pp\to\tw_i\overline{\tw}_j X$ (three reactions), and
$pp\to\tz_i\tz_jX$ (ten reactions).  
In Fig.~1, we illustrate total pair production cross sections at 
$\sqrt{s}=14$ TeV (LHC energy) for 
$\tw_1\tz_1$, $\tw_1\tz_2$, $\tw_1\overline{\tw}_1$, and
$\tz_2\tz_2$ production. We have convoluted with EHLQ Set 1
parton distributions\cite{EHLQ}. We show curves versus $m_{\tg}$ for
($m_{\tq}/m_{\tg},\tan\beta)$ =\hbox{(1,2)} (solid), 
(1,20) (dashed) and (2,2) (dotted), and have taken $\mu =-m_{\tg}$
throughout. For $m_{\tg}$ on the low end of the scale, cross sections can
be very large due to production via {\it on-shell} $W$ and $Z$ decays.
For larger values of $m_{\tg}$, the chargino and neutralino masses
increase, and the cross sections decrease rapidly because production now
takes place via {\it off-shell} $W$, $\gamma$ 
and $Z$ graphs, as well as 
squark exchange. Even so, we see that the $\tw_1\overline{\tw}_1$ and
$\tw_1\tz_2$ 
cross sections remain above the 0.1 $pb$ level, owing to a large
gauge coupling, even for $m_{\tg}\sim 1000$ GeV. 
Other chargino-neutralino production processes occur
at typically much lower rates, and hence are less likely to give
interesting phenomenology. The most interesting of the chargino-neutralino 
reactions, as we shall see,
is $\tw_1\tz_2$ production. This production rate is actually highest for large
values of $m_{\tq}$, due to negative interference between squark exchange
and $W^*$ exchange amplitudes. 

The branching fractions for two 
decay modes of the light chargino, $\tw_1$, are shown 
versus $m_{\tg}$ in Fig.~2, again for {\it a}) 
$(m_{\tq}/m_{\tg},\tan\beta)=\hbox{(1,2)}$, {\it b}) (1,20) and {\it c}) (2,2),
with $\mu =-m_{\tg}$. The dashed curves shows the branching fraction for
$\tw_1^+\to\mu^+\nu_{\mu}\tz_1$, which typically varies between 10-20\%,
depending on parameter choices, and is $\sim 11\%$ for large values of
$m_{\tq}$ and $m_{\tell}$, for which decay via virtual $W$ dominates. For 
small values of $m_{\tg}$, the decay $\tw_1\to\ell\tnu$ is kinematically 
accessible, and 
is the dominant decay mode. For
values of $m_{\tg}> 550\hbox{--}600$ GeV, two body decays to real $W$ bosons
become kinematically allowed, and dominate the $\tw_1$ branching fractions
in this region.

Several decay modes of the neutralino $\tz_2$ are shown 
versus $m_{\tg}$ in Fig.~3, for the same cases {\it a}), {\it b}) and
{\it c}) as in Fig.~2.  
The dashed curves shows the branching fraction for
$\tz_2\to\mu^+\mu^-\tz_1$, which is $\sim 10-20\%$ for $m_{\tq}\sim m_{\tg}$, 
but only a few per cent for $m_{\tq}\sim 2m_{\tg}$, where decay via a virtual
$Z$ becomes important. Two body modes such as $\tz_2\to\nu\tnu$ dominate
for small $m_{\tg}$. Just as for the light 
chargino, other two body decay modes of $\tz_2$ open up around $m_{\tg}\sim 600$ GeV.
In case {\it a}), the decay $\tz_2\to\tz_1 H_{\ell}$ dominates, and one
expects very few leptons from $\tz_2$. In case {\it b}) and {\it c}), the
mode $\tz_2\to\tz_1 Z$ opens up first, and there is a region in which one
expects real $Z\to\ell\bar{\ell}$ in the event sample.
As we shall see, the opening of these two body $\tz_2$ modes can effectively 
spoil the clean trilepton signal from $\tw_1\tz_2$ production, in
one case because the leptonic branching fraction for the $Z$ is rather
small, and additional background from $WZ$ appears, 
and in the other because the Higgs boson essentially
always decays to $b$-quarks.

In order to assess detection prospects for charginos and neutralinos at LHC
energy, we use the event simulation program ISAJET 7.07\cite{ISAJET}.
For a given input parameter set,
$m_{\tg}, m_{\tq},\mu ,\tan\beta ,m_{H_p}, m_t$
and $m_{\tell_L},m_{\tell_R},m_{\tnu_L}$,
(recall that $m_{\tell_L},m_{\tell_R},m_{\tnu_L}$ are determined by
$m_{\tg}, m_{\tq}$ and $\tan\beta$)
the routine ISASUSY calculates all sparticle masses and branching fractions
to various decay modes. ISAJET then produces all combinations of
chargino and neutralino production subprocesses,
in proportion to their respective cross sections.
The charginos and neutralinos then decay via the various 
cascades with appropriate
branching fractions as given by the MSSM. Radiation of 
initial and final state
partons is also included in ISAJET. Final state quarks and gluons are 
hadronized, and
unstable particles are decayed until stable final states are reached.
Underlying event activity is also modeled in ISAJET.

For event simulation at the LHC, 
we use the toy calorimeter simulation package ISAPLT.
We simulate calorimetry with cell size 
$\Delta\eta\times\Delta\phi =0.05\times 0.05$, which extends between
$-5.5<\eta <5.5$. We take hadronic 
energy resolution to be $50\% /\sqrt{E_T}$ for $|\eta |<3$, and to be
a constant 
$10\%$ for $3<|\eta |<5.5$, to model the effective $p_T$ resolution of
the forward calorimeter including the effects of shower spreading.
We take electromagnetic resolution to be $15\% /\sqrt{E_T}$. 
Jets are coalesced
within cones of $R=\sqrt{\Delta\eta^2 +\Delta\phi^2} =0.7$ using
the ISAJET routine GETJET. For the purpose of jet veto (essential to 
eliminate top quark background),
clusters with $E_T>25$ GeV 
are labeled as jets.
Muons and electrons are classified as isolated if they have $p_T>10$ GeV,
$|\eta (\ell )|<2.5$,
and the visible activity within a cone of $R =0.3$ about the lepton 
direction is less than $E_T({\rm cone})=5$ GeV.

\section{Signal and background}

In our simulation of chargino and neutralino events, we generate all
twenty-one of the reactions referred to in Sec. II. We first classify 
signals according to the number of isolated leptons present in the
signal events. We found observable 
signal cross sections in the $0\ell$, $1\ell$,
$2\ell$ and $3\ell$ channels. However, the $0\ell + {\rm jets} +\eslt$ sample should
be dominated by other sources of SUSY events, such as gluino and squark 
production, as well as SM backgrounds. The $1\ell +{\rm jets}+\eslt$ channel
yielded observable signal rates; however, these were dominated by large
SM backgrounds from $W\to\ell\nu_{\ell}$ ($\ell =e\ or\ \mu$), 
$W\to\tau\nu_{\tau}$ and $t\bar t$ production, as well as single lepton
events from gluino and squark production\cite{BTW}. Likewise, the opposite sign (OS)
dilepton ($2\ell +{\rm jets}+\eslt$) sample, which has a substantial 
rate due especially to $\tw_1\overline{\tw}_1$ production, suffers large 
SM backgrounds mainly from $t\bar t$ production as well as other SUSY 
sources\cite{BTW}. Same-sign (SS) dilepton events can occur from processes such as
$\tw_1\tz_2$ production, where one of the decay leptons is soft or missed
through a crack in the detector, but these rates are small compared to SS
dilepton production from squarks and gluinos\cite{BTW,ALSO}. Unlike these events,
the $\tw_1\tz_2$ events would usually be free from jet activity. In this paper,
we focus on the zero-jet (clean) trilepton signal, 
which occurs at a substantial rate due to $\tw_1\tz_2$ production, and which,
with an appropriate set of cuts, is relatively free of SM backgrounds. There is
also a possibility for $4\ell$ events from sources such as $\tz_2\tz_2$
production followed by subsequent leptonic decays. These signals have been
considered (as backgrounds to Higgs boson decay to neutralino pairs) in Ref.
\cite{BBKT}; cross sections range up to a few $fb$
after cuts, so the signal is not
large, although SM physics backgrounds can be eliminated. In addition, $5\ell$ signals
have been considered in Ref. \cite{BARB}; we did not find significant rates for
signal in this channel. 

To assess the viability of the trilepton signal at LHC energy, we 
use ISAJET 7.07 to generate $\sim 100K$ events for the following four
cases, where $m_{\tq}=m_{\tg}+20$ GeV, and $\tan\beta =2$:
%
%
\def\unstupidspace{\setbox0=\hbox{ }\hskip-\wd0\relax}
\begin{description}
\item[\rm\hbox{\hskip20pt}\unstupidspace Case~1: ] $m_{\tg} =-\mu =300$ GeV,
$m_{\tw_1}=95.8$ GeV, $m_{\tz_2}=96.4$ GeV, $m_{\tz_1}=45$ GeV,
\item[\rm\rm\hbox{\hskip20pt}\unstupidspace Case~2: ] $m_{\tg} =-\mu =400$ GeV,
$m_{\tw_1}=123.5$ GeV, $m_{\tz_2}=123.8$ GeV, $m_{\tz_1}=59.8$ GeV,
\item[\rm\rm\hbox{\hskip20pt}\unstupidspace Case~3: ] $m_{\tg} =-\mu =500$ GeV,
$m_{\tw_1}=152.6$ GeV, $m_{\tz_2}=152.8$ GeV, $m_{\tz_1}=74.8$ GeV,
\item[\rm\rm\hbox{\hskip20pt}\unstupidspace Case~4: ] $m_{\tg} =-\mu =600$ GeV,
$m_{\tw_1}=182.7$ GeV, $m_{\tz_2}=182.8$ GeV, $m_{\tz_1}=90.0$ GeV.
\end{description}
The above parameters are motivated by predictions from supergravity GUT
models with radiative electroweak symmetry breaking\cite{AN,SUGRA}. 
We also generate SM background event samples from $WZ$ production, as well
as from $t\bar t$ production for $m_t=135$ GeV and $175$ GeV. Finally, we 
generate samples with all other possible SUSY particle production processes,
to see if the $\tw_1\tz_2$ component can be isolated.

We first select events by requiring
\begin{itemize}
\item {\it three} isolated leptons; the two fastest have 
$p_T(\ell_1,\ell_2 )>20$ GeV, 
while the third has $p_T(\ell_3 )>10$ GeV.
\end{itemize}
The signal cross section before and after cuts for the above four cases, SM 
backgrounds, and trilepton cross sections from other
SUSY sources, are listed in Table 1. At this point, the signal can still
be dominated by SM background, but even more so by trilepton events from 
gluino and squark production. Gluino and squark events 
should contain substantial jet activity as well as a hard $\eslt$
spectrum. We illustrate the latter for cases 1-3 in Fig.~4: the $\eslt$
spectrum is clearly harder for $\tg$ and $\tq$ events. Hence, we require 
in addition
\begin{itemize}
\item no central jets, ($p_T({\rm jet})>25$ GeV; $|\eta (jet)|<3$), and
\item $\eslt <100$ GeV.
\end{itemize}
Contributions to the trilepton signal from $\tg$ and $\tq$ production are 
now smaller than the chargino-neutralino signal.
At this point, the dominant background is from $WZ\to 3\ell$ production,
so we require
\begin{itemize}
\item for OS same flavor dileptons, $m(\ell\bar{\ell})\ne M_Z\pm 8$ GeV.
\end{itemize}
This reduces the $WZ$ background to below the $fb$ level, but leaves a
significant $t\bar t$ background, especially for the $m_t=135$ GeV 
case\cite{BARB}.
The latter background can be further reduced by splitting the event 
sample in two. In the first, we require:
\begin{itemize}
\item the two fastest leptons be same sign (SS); and the flavor of the 
slow lepton be the same as
(but anti-) the flavor of either of the two fast leptons.
\end{itemize}
This diminishes the signal by a factor of about 2, but almost completely removes
top quark background, from which the two hardest leptons, almost always, come
from the primary decays of the $t$-quarks, and hence have opposite signs.
Some of the rejected signal can be recovered by also accepting events with:
\begin{itemize}
\item two fastest leptons of opposite sign (OS) if $p_T({\rm slow\
lepton})>20$ GeV,
\end{itemize}
which is more effective in reducing top quark background than signal.

The sum of these two classes of cuts are listed in the last row of Table~I,
where we find signal in the 13--40~$fb$ range, with SM background at the
level of 0.5--3~$fb$, depending on $m_t$. In addition, there exists an
irremovable contribution, shown in parenthesis, 
from other SUSY sources. We investigated this
remaining SUSY background, and found it to be all either associated
production events ({\it e.g.} $\tg\tz_2$, etc.,) or slepton pair events:
the gluino and squark pairs had been completely eliminated. 
For case~4, the spoiler mode $\tz_2\to\tz_1 H_{\ell}$ has opened up, thus
destroying the trilepton signal.

In Table~II, we list the signal cross sections after all the above cuts
as a matrix in $m_{\tg}$ vs. $\mu$ for $\tan\beta = 2$. 
Starred entries are in the LEP
excluded region. We see that signal cross sections are 
usually larger for negative $\mu$ than for positive $\mu$, due mainly to
a larger $\tz_2\to\tz_1\ell\bar{\ell}$ branching ratio\cite{BT}.
Also, for negative $\mu$, the signal is killed much beyond $m_{\tg}=500$ GeV,
while for positive $\mu$ one gets a robust signal past $m_{\tg}=600$ GeV,
especially for the supergravity favored choice,
$\mu =m_{\tg}$. For $\mu =\pm 100$ GeV, a signal of a 
few $fb$ persists out to $m_{\tg}\sim 700\hbox{--}900$ GeV. In fact, for this
region of parameter space, most of the trileptons come from subprocesses other 
than $\tw_1\tz_2$, with {\it e.g.} $\tw_1\tz_3$, $\tw_2\tz_4$, etc.,
also contributing.

How does the signal depend on other choices of $m_{\tq}$ and $\tan\beta$?
We show in Fig.~5 data points for signal rates for $\mu =\pm m_{\tg}$, for
{\it a}) the case already considered, 
($m_{\tq}/m_{\tg},\tan\beta {\rm )}=(1,2)$,
and also {\it b}) (2,2), {\it c}) (1,20) and finally {\it d}) (2,20).
SM background level is indicated by the dotted line, for the worst case
with $m_t=135$ GeV, and the dashed line for $m_t=175$ GeV.
We see that frames {\it a}) and {\it c}), with $m_{\tq}\sim m_{\tg}$ so that
$m_{\tell} \ll m_{\tq}$, yield the largest signal cross sections, and these
signals remain substantially above background out to $m_{\tg}\sim
500\hbox{--}600$ GeV.
Overall behavior is similar for both large and small values of $\tan\beta$.
For frames {\it b}) and {\it d}), with $m_{\tq}=2m_{\tg}$ so that 
sleptons are quite heavy, signal rates drop to the several $fb$ level, which
may be observable above background for the case of a heavier top quark.

As can be seen in Fig.~3, the two body decay $\tz_2\to\tz_1Z$ opens up
and is dominant for frame {\it b}) and {\it c}) for $m_{\tg}\sim
600\hbox{--}800$ GeV.
In this case, our cut of $m(\ell\bar{\ell})\ne M_Z\pm 8$ GeV will 
also eliminate the signal. To see if this signal can still be gleaned from 
background, we implement all the above cuts except the offending $Z$
mass cut. We set $m_{\tg}=700$ GeV, and take $m_{\tq}=2m_{\tg}$, with 
$\tan\beta =2$. The trilepton cross section is then at the 3.1~$fb$ level, 
while $WZ$ background is 59~$fb$. We attempt to remove the $WZ$
background by requiring 
transverse mass 
$m_T(\ell,\eslt )>100$ GeV, to exclude the real $W$ Jacobian peak. This 
reduces both signal (which is already small on account of the 
rather large $\tw_1$ and $\tz_2$ masses) and background to the 1~$fb$ level,
making distinction of this tiny signal very difficult. Hence, the
$\tz_2\to\tz_1 Z$ two body decay also acts as a spoiler mode for trilepton
events, in part because the signal becomes rate-limited due 
to the small leptonic
branching ratio of the $Z$ boson together with the fact that
the the real $Z$ mode is open only when the $\tz_2$ (and $\tw_1$)
is rather heavy, 
but also due to irremovable $WZ$ background. 

In addition, we have investigated whether the $\tw_1\tz_2$ signal with
$\tw_1\to\tz_1 W$ ($W \to \ell\nu$) and $\tz_2\to\tz_1 H_{\ell}$
($H_{\ell}\to b\bar b$) is observable. Here, 
we looked for a single lepton plus two central jet signal.
In addition, we required one $b$-jet 
to have its decay vertex tagged with an efficiency as given in Ref.~\cite{BKT}.
We then looked for a mass bump at $m_{jj}= m_{H_{\ell}}$. The mass
bump was unfortunately obscured by a $t\bar t$ background $\sim 100$
times greater 
than signal. Hence, we affirm that the $\tz_2\to\tz_1 H_{\ell}$
decay mode is indeed a spoiler.

\section{Constraining chargino and neutralino masses}

For the values of parameters examined in this paper, there will be a plethora
of various signals from SUSY at the LHC just from gluino and squark
production\cite{BTW,ALSO}. The unique feature of the trilepton signal for
LHC is that it offers the possibility of reasonably clean information
on sparticle masses from which to start to unravel the whole 
SUSY particle spectrum.
With a relatively pure sample of signal events, and
an event structure consisting of only three isolated leptons plus no jets,
it is especially easy to reconstruct where each lepton came from: in
$\ell\ell'\bar\ell'$ events, the $\ell'\bar\ell'$ come from the neutralino,
whereas in $\ell\ell\bar{\ell}$ events, the OS dilepton with smallest 
transverse opening angle usually comes from the neutralino. 

We show in Fig.~6 the invariant mass $m(\ell\bar{\ell})$ in trilepton events
after all cuts, for cases 1, 2 and 3 above. We show the SM background (dots)
and the SUSY signal plus SM and SUSY background (solid), where we have taken the
mass of the same flavor OS dilepton pair in $\ell\ell'\bar\ell'$ events,
and the mass of the OS pair with the smaller transverse opening angle in
$\ell\ell\bar{\ell}$ events. Kinematically, the mass spectrum is constrained 
to lie between $0<m(\ell\bar{\ell})<m_{\tz_2}-m_{\tz_1}$. The sharp
cutoff at the upper end-point is evident from these plots, offering
a clean measure of $m_{\tz_2}-m_{\tz_1}$. The corresponding value of
$m_{\tz_2}-m_{\tz_1}$ in these figures is 51, 64 and 78 GeV, for cases
1, 2 and 3, respectively. In models with $|\mu|$ much
larger than the electroweak gaugino masses (as is the case
in SUGRA models), one frequently expects
$2m_{\tz_1}\simeq m_{\tz_2}\simeq m_{\tw_1}\simeq {1\over 3}m_{\tg}$
\cite{MSSM}, so that the
cutoff in $m(\ell\bar{\ell})$ occurs at approximately $m_{\tz_1}$: this is true
in our simulations.

Are there other distributions which can yield significant information
constraining sparticle masses? Explicit mass reconstruction is, of course,
not possible since in each event one is missing a neutrino and the two massive
$\tz_1$ particles. However, we examined a variety of distributions, 
including trilepton invariant mass, $m(3\ell )$, the summed scalar transverse
energy, $\Sigma E_T =p_T(\ell_1 )+p_T(\ell_2)+p_T(\ell_3)+\eslt$, and the
three lepton plus missing energy cluster transverse mass, $m_T(3\ell ,\eslt)$. 
All these distributions suffered substantial smearing due mainly to the
continuum nature of the underlying $2\to 2$ subprocess; additional
smearing is expected for distributions involving $\eslt$ if LHC is run at 
high luminosity, where event pile-up becomes a problem. The distributions
do scale with overall sparticle mass values. An example is given in Fig. 7
for the cluster transverse mass. The solid histograms show the 
$m_T(3\ell,\eslt )$ distribution for cases 1, 2 and 3, after all cuts. The 
distribution maxima increases with increasing sparticle mass, as does
the distribution mean, which is 154, 175 and 195 GeV for the respective cases.
The distributions are also sensitive to the $\tz_1$ mass, in that if 
$m_{\tz_1}\to 0$, there is more energy available to make visible decay
products. To illustrate this, we show via dashed histograms the
distribution shapes where we have by hand set the $\tz_1$ mass
to zero without changing other masses and branching fractions. In this case,
the distribution maxima move substantially to higher energy, and the means move
to 195, 225 and 271 GeV, respectively. If a trilepton signal is found above
expected background levels, then the shapes of distributions such as cluster
transverse mass or summed scalar $E_T$ will also serve to constrain the 
sparticle masses. For example, for the plots in Fig. 7, the distribution means
can be parameterized as 
\begin{eqnarray}
\langle m_T(3\ell,\eslt ) \rangle = 0.69\times (m_{\tw_1}+m_{\tz_2})
-0.95\times m_{\tz_1}+60\ {\rm GeV}.
\eqnum{3}
\end{eqnarray}
For the various cases in Fig. 7, this reproduces the distribution means
to within 3-6 GeV. We have also checked that this is an adequate fit 
for cases where just the chargino
mass is reduced by 20 GeV. It should, however, be remembered that the
fit in Eq. (3) is sensitive to the details of the cuts and detector simulation.
Our purpose in showing this is to illustrate that it should be possible
to obtain further information on the sparticle masses from the transverse mass,
summed scalar $E_T$, and trilepton invariant mass  distributions. 

In addition, we have investigated a variety of other distributions which show
sensitivity to sparticle masses. In $\ell'\ell\bar{\ell}$ events, the
transverse opening angle $\Delta\phi (\ell\bar{\ell})$ decreases with increasing
$m_{\tz_2}\over m_{\tz_1}$, for fixed $m_{\tz_2}-m_{\tz_1}$: {\it i.e.} the 
dilepton pair becomes more tightly collimated. Also, the opening angle
$\Delta\phi (\ell ',\ell\bar{\ell})$ (between $\ell '$ and the vector sum of 
$\ell$ and $\bar\ell$) increases with increasing $m_{\tw_1}\over m_{\tz_1}$
(again, for fixed $m_{\tz_2}-m_{\tz_1}$), so that events are more nearly back-to-back.
Moreover, the $p_T(\ell ')$ distribution is sensitive to $m_{\tw_1}$. 
Clearly, if a signal is observed, a variety of distributions will have to be
tested against various sparticle mass hypotheses. Likelihood functions can then
be constructed to ascertain the most probable sparticle mass combination. 


\section{Summary}

We have reexamined the signal from the production of charginos and neutralinos
at the LHC using ISAJET 7.07, incorporating experimental
conditions corresponding to a generic LHC 
detector. As in Ref.\cite{BARB}, we find that the
reaction $p\bar{p} \to \tw_1\tz_2 \to \ell\bar{\ell}\ell'$ provides
the best prospects for the identification of the signal. The signal
thus consists of events with three hard, isolated leptons and essentially
no jet activity. We have devised a set of cuts to reduce backgrounds 
from top quarks and $WZ$ production to negligible levels {\it provided}
that two body decays of charginos and neutralinos are kinematically
inaccessible. The observation of this signal would be
direct evidence for neutralino production; this is especially
important since the production
of (gaugino-like) neutralinos by $e^+e^-$ collisions is strongly suppressed
unless the selectron is also rather light.
The effect of the various cuts as well as the signal 
level for representative choices of parameters is shown in Table I.
We also mention that with these cuts, other SUSY sources of
trileptons such as squark or gluino pair production, or the
production of gluinos and squarks in association with a chargino
contribute between just 3--15\% to the signal. The relatively
clean sample of chargino and neutralino events, as we will see,
enables us to obtain experimental constraints on their masses.

We see from Table I that the signal cross section exceeds 10 $fb$
(corresponding to more than 100 events per year even assuming the lower value
for the LHC design luminosity) for chargino and neutralino masses up to about
150 GeV, corresponding to $m_{\tg} \simeq 500$ GeV. For yet heavier sparticles
(case 4 in Table I), the decays $\tz_2 \to \tz_1 H_{\ell}$,
$\tz_2 \to \tz_1Z$ become kinematically
accessible. These then dominate the decays of $\tz_2$.
In the first case, the trilepton cross
section is reduced to an unobservable level since the Higgs boson
decays to heavy fermions. In the other case where two leptons
come from the decay of a real Z, there remains a 
background from $WZ$ production; although a signal to
background ratio of 1:1 is possible after a transverse mass
cut, the signal appears to be too small for this strategy
to be viable. The dependence of the 
signal on the superpotential parameter $\mu$ is shown in Table II, while the 
variation with $\tan\beta$ and $m_{\tq}$ is illustrated in Fig. 5.

Motivated by the fact that after our cuts we are left with a relatively
uncontaminated sample of $\tw_1\tz_2$ events, we have examined the prospects
for measuring chargino and neutralino masses from these events. Since 
the invariant mass of dileptons from $\tz_2$ decays
is kinematically constrained to be smaller than $m_{\tz_2}-m_{\tz_1}$,
this mass difference can be inferred from the upper edge of the distribution 
shown in Fig. 6. Since at least several hundred trilepton events are expected
at the LHC with an integrated luminosity of 10-20 $fb^{-1}$, 
it should be possible to construct this distribution
rather well. Unfortunately, 
it is not possible to directly reconstruct the masses of
the $\tw_1$ or $\tz_2$ because 
two neutralinos and the neutrino are undetected in
every event. We have shown, however, that by studying the shapes
of other
distributions such as the $m_T(3\ell,\eslt)$-distribution (see Fig. 7), or 
the $\Sigma E_T$-distribution, whose means may be expected
to scale with parent masses as discussed in Sec. IV, it should be possible
to obtain one further constraint between $m_{\tz_1}$, $m_{\tz_2}$ and 
$m_{\tw_1}$. Ultimately, matching a variety of observed distributions
against different sparticle mass hypotheses should allow the most probable
combination of sparticle masses consistent with data to be selected.
These experimental constraints
may serve as a relatively clean starting point for the procedure of unraveling
the whole spectrum of SUSY particle masses. Such information ought to help
test the ideas behind supergravity grand unification (for instance, the 
unification of gaugino masses), and further,
to aid in sorting out more complex events from the cascade decays of 
gluinos and squarks.

%
\acknowledgments

This research was supported in part by the U.~S. Department of Energy
under contract number DE-FG05-87ER40319, DE-AC02-76CH00016, and
DE-AM03-76SF00235. In addition, the work of HB was supported by the
TNRLC SSC Fellowship program. 
%
%
%
%

%
\newpage
%
%

\begin{table}
\caption[]{Cross sections (in $fb$) 
after cuts for chargino-neutralino production at LHC for cases~1--4 listed
in Sec. III of the text, along with SM backgrounds. 
Contributions from other SUSY particles are listed in parenthesis below
signal rates. We take $\mu =-m_{\tg}$ and $\tan\beta =2$.
The notations $3\ell$, $0j$, $\eslt$, and $M_Z$ refer to the
trilepton, jet veto, missing energy, and $Z$-mass veto cuts described
in the text. The ``$SS,FL$'' subsample has the two fastest leptons
with the same sign and the third with the opposite flavor; the
``$OS,L20$'' sample has the two fastest leptons with opposite signs
and $p_T(\hbox{slow lepton})> 20$~GeV. Results have 
summed over $e$'s and $\mu$'s. We do not show other SUSY contributions to case 4
because we do not consider this signal to be observable.}
\bigskip
\begin{tabular}{lccccccc}
cuts & case 1 & case 2 & case 3 & case 4 & $t\bar t(135)$ & $t\bar t(175)$ & 
$WZ$ \\
\tableline
\tableline
{\it none} & $\ds{11.1K\atop (1521K)}$ & $\ds{5.1K\atop (386K)}$ & 
$\ds{2.6K\atop (124K)}$ & 1.5K & 2611K & 841K & 17K \ts\\
\tableline
$3\ell$ & $\ds{151\atop (17.5K)}$ & $\ds{95\atop (4.6K)}$ &
$\ds{51\atop (1.3K)}$ & 1.6 & 486 & 98 & 117 \ts\\
\tableline
$3\ell ,0j$ & $\ds{73\atop (15)}$ & $\ds{42\atop (3.9)}$ &
$\ds{22\atop (1.0)}$ & 0.6 & 50 & 2.5 & 63 \ts\\
\tableline
$3\ell ,0j,\eslt$ & $\ds{72\atop (8.4)}$ & $\ds{40\atop (2.4)}$ &
$\ds{20\atop (0.4)}$ & 0.6 & 48 & 2.1 & 59 \ts\\
\tableline
$3\ell ,0j,\eslt ,M_Z$ & $\ds{67.5\atop (8.4)}$ & $\ds{37.4\atop (2.4)}$ & 
$\ds{19\atop (0.4)}$ & 0.4 & 42 & 1.9 & 0.8 \ts\\
\tableline
\tableline
$\ds{3\ell ,0j,\eslt ,M_Z\atop SS,FL}$ & $\ds{26\atop (2.1)}$ &
$\ds{15\atop (1.8)}$ & $\ds{7.1\atop (0.2)}$ & 0.3 & 0.4 & $<0.2$ & 0.3 \ts\\
\tableline
$\ds{3\ell ,0j,\eslt ,M_Z\atop OS,L20}$ & $\ds{15\atop (3.5)}$ &
$\ds{9.0\atop (0.3)}$ & $\ds{6.1\atop (0.1)}$ & 0.1 & 2.0 & $<0.2$ & 0.2 \ts\\
\tableline
\tableline
$\ds{3\ell ,0j,\eslt ,M_Z\atop SS,FL\ or\ OS,L20}$ & $\ds{41\atop (5.6)}$ & 
$\ds{24\atop (2.1)}$ & $\ds{13\atop (0.3)}$ & 0.4 & 2.4 & $<0.2$ & 0.5 \ts\\
\end{tabular}
\end{table}

\iftightenlines\else\newpage\fi

\begin{table}
\caption[]{Trilepton cross sections in $fb$ after all cuts for various values 
of $m_{\tg}$ and $\mu$. A star indicates a point excluded by LEP. We take
$m_{\tq} =m_{\tg}+20$ GeV, and $\tan\beta =2$. The total SM background is
2.9~$fb$ for $m_t =135$~GeV, and $<0.7$ $fb$ for $m_t =175$ GeV.  
Cuts are described in the text. Results have summed over $e$'s and $\mu$'s. }

\bigskip

\begin{tabular}{crrrrrrrrrr}
$m_{\tg}\backslash\mu$ & $-m_{\tg}$ & $-400$ & $-300$ & $-200$ & $-100$ & 
100 & 200 & 300 & 400 & $m_{\tg}$ \\
\tableline
300 & 41.5 & 49.6 & 41.5 & 25.9 & 1.7 & * & 10.1 & 23.0 & 20.9 & 23.0 \\
400 & 23.7 & 23.7 & 18.7 & 3.9 & 2.2 & * & 4.3 & 10.6 & 14.7 & 14.7 \\
500 & 13.2 & 8.7 & 4.3 & 1.3 & 1.0 & 4.1 & 0.9 & 1.9 & 7.3 & 11.5 \\
600 & 0.4 & 0.2 & 0.1 & 0.8 & 1.1 & 4.6 & 1.1 & 0.4 & 1.8 & 8.8 \\
700 & 0.1 & 0.0 & 0.1 & 0.6 & 1.2 & 2.7 & 0.7 & 0.2 & 0.8 & 0.0 \\
900 & 0.0 & 0.0 & 0.0 & 1.3 & 0.2 & 1.9 & 0.1 & 0.4 & 0.0 & 0.0 \\
\end{tabular}
\end{table}


%
\begin{figure}
\caption[]{Total cross section for {\it a}) $\tw_1\tz_1$, {\it b}) $\tw_1\tz_2$,
{\it c}) $\tw_1\overline{\tw}_1$ and {\it d}) $\tz_2\tz_2$ production in $pp$
collisions at $\sqrt{s}=14$ TeV. Curves are for 
$(m_{\tq}/m_{\tg},\tan\beta )=\hbox{(1,2)}$ (solid), (1,20) (dashes), and
(2,2) (dots). We have taken $\mu =-m_{\tg}$.}
\end{figure}
%
\begin{figure}
\caption[]{Selected branching fractions for $\tw_1$ decay versus $m_{\tg}$,
for
{\it a}) $(m_{\tq}/m_{\tg},\tan\beta )=\hbox{(1,2)}$, {\it b}) (1,20), and
{\it c}) (2,2). We have taken $\mu =-m_{\tg}$. The dashed curve is for
$\tw_1\to\mu\nu_{\mu}\tz_1$, while solid is for $\tw_1\to\tz_1 W$.}
\end{figure}
%
\begin{figure}
\caption[]{Selected branching fractions for $\tz_2$ decay versus $m_{\tg}$,
for
{\it a}) $(m_{\tq}/m_{\tg},\tan\beta )=\hbox{(1,2)}$, {\it b}) (1,20), and
{\it c}) (2,2). We have taken $\mu =-m_{\tg}$. The dashed curve is for
$\tz_2\to\mu\bar{\mu}\tz_1$, while solid is for $\tz_2\to\tz_1 H_{\ell}$, 
and dotted is for $\tz_2\to\tz_1 Z$.}
\end{figure}
%
\begin{figure}
\caption[]{Distribution in missing transverse energy ($\eslt$) at LHC from
{\it a}) all chargino-neutralino events and {\it b}) all other supersymmetric
sources, after requiring three isolated leptons. 
We have illustrated spectra for text cases 1, 2 and 3
corresponding to $m_{\tg}=300,\ 400$ and 500 GeV.}
\end{figure}
%
\begin{figure}
\caption[]{Total cross section for trilepton signal after all cuts in the text,
versus $m_{\tg}$, 
for {\it a}) $(m_{\tq}/m_{\tg},\tan\beta )=\hbox{(1,2)}$, {\it b}) (2,2),
{\it c}) (1,20) and {\it d}) (2,20). We plot for $\mu=-m_{\tg}$ (x's)
and $\mu=+m_{\tg}$ (o's). The dotted line corresponds to the SM background
expected of $WZ$ and $t\bar t(135)$ while the dashed line denotes the
same background for $m_t = 175$ GeV}
\end{figure}
%
\begin{figure}
\caption[]{Distribution in OS dilepton invariant mass from both SUSY
and SM sources (for $m_t=175$ GeV) after all cuts given in the text, for
{\it a}) case 1, {\it b}) case 2 and {\it c}) case 3,
corresponding to $m_{\tg}=300,\ 400$ and 500 GeV. For $\ell\ell'\bar{\ell}'$
events, we plot the mass of the same-flavor pair, while for 
$\ell\ell\bar{\ell} $, we plot the mass of the OS pair with smallest 
transverse opening angle.}
\end{figure}
%
\begin{figure}
\caption[]{Distribution in trilepton plus $\eslt$ cluster transverse
mass from both SUSY and SM sources after all cuts given in the text, for
{\it a}) case 1, {\it b}) case 2 and {\it c}) case 3,
corresponding to $m_{\tg}=300$, 400 and 500 GeV (solid histogram).
The dashed histograms are for corresponding distributions after setting
$m_{\tz_1}=0$ by hand.}
\end{figure}
\vfill\eject
\centerline{\epsfbox{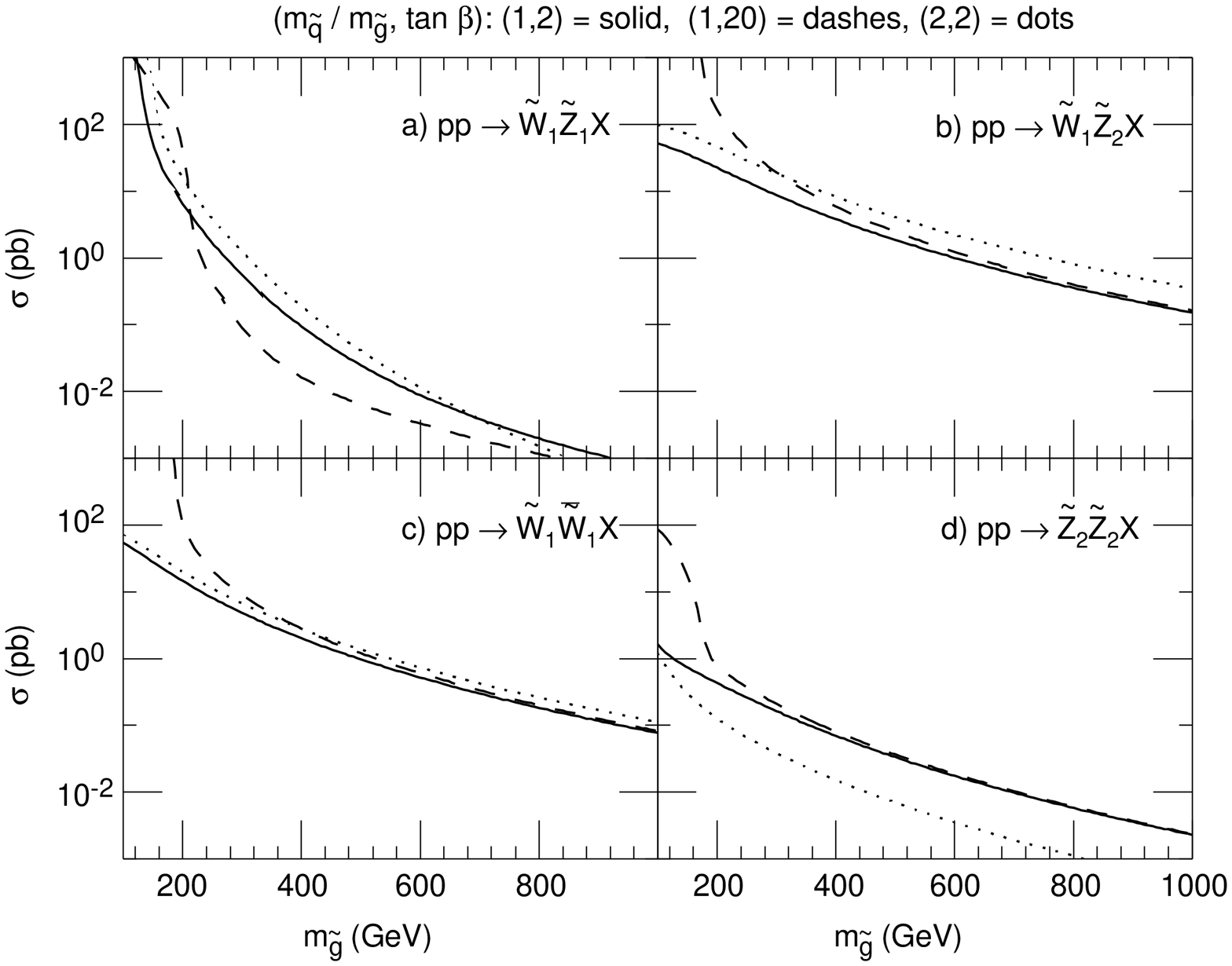}}
\bigskip\bigskip
\centerline{FIG.~1}
\vfill\eject

\centerline{\epsfbox{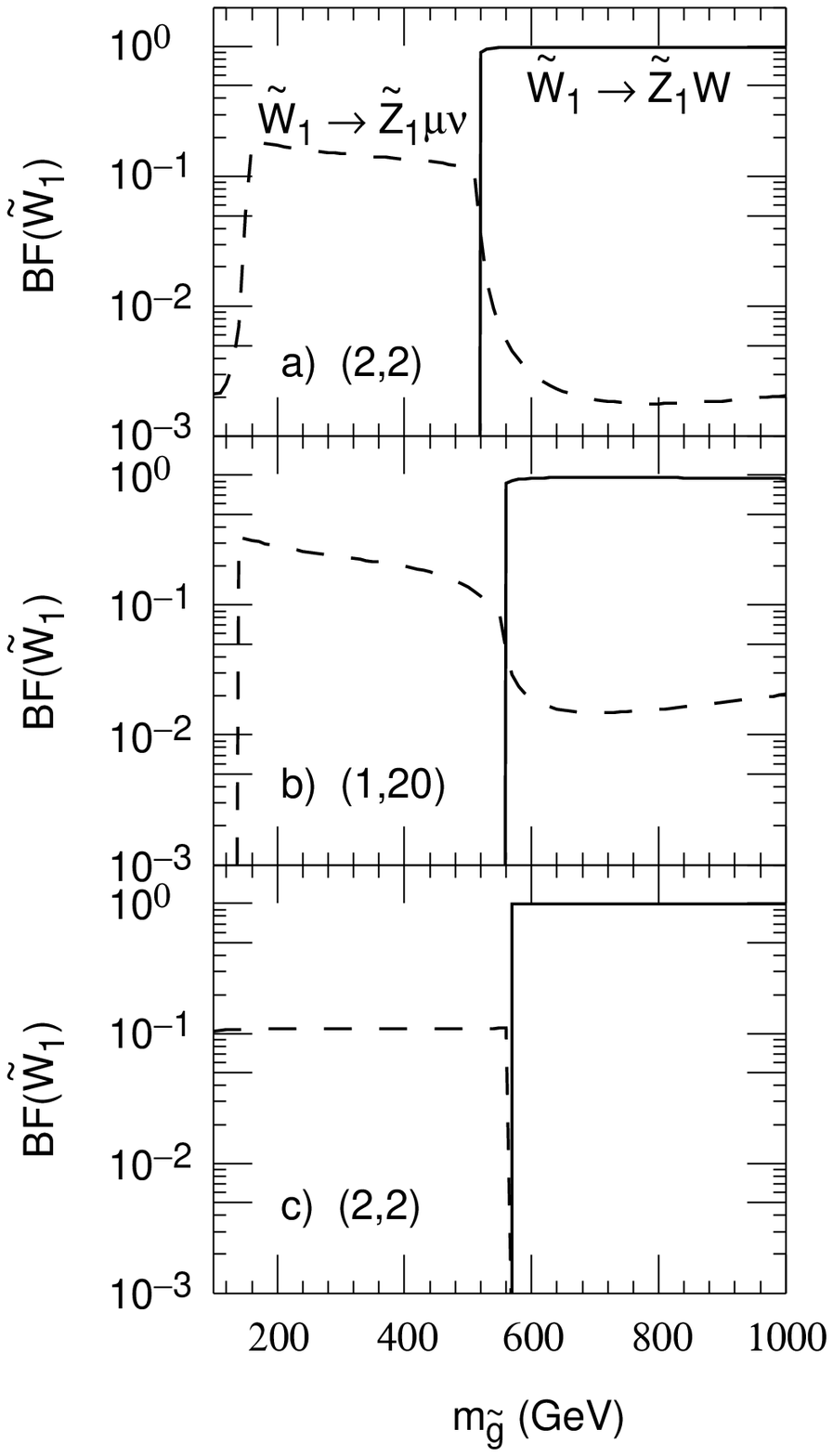}}
\bigskip\bigskip
\centerline{FIG.~2}
\vfill\eject

\centerline{\epsfbox{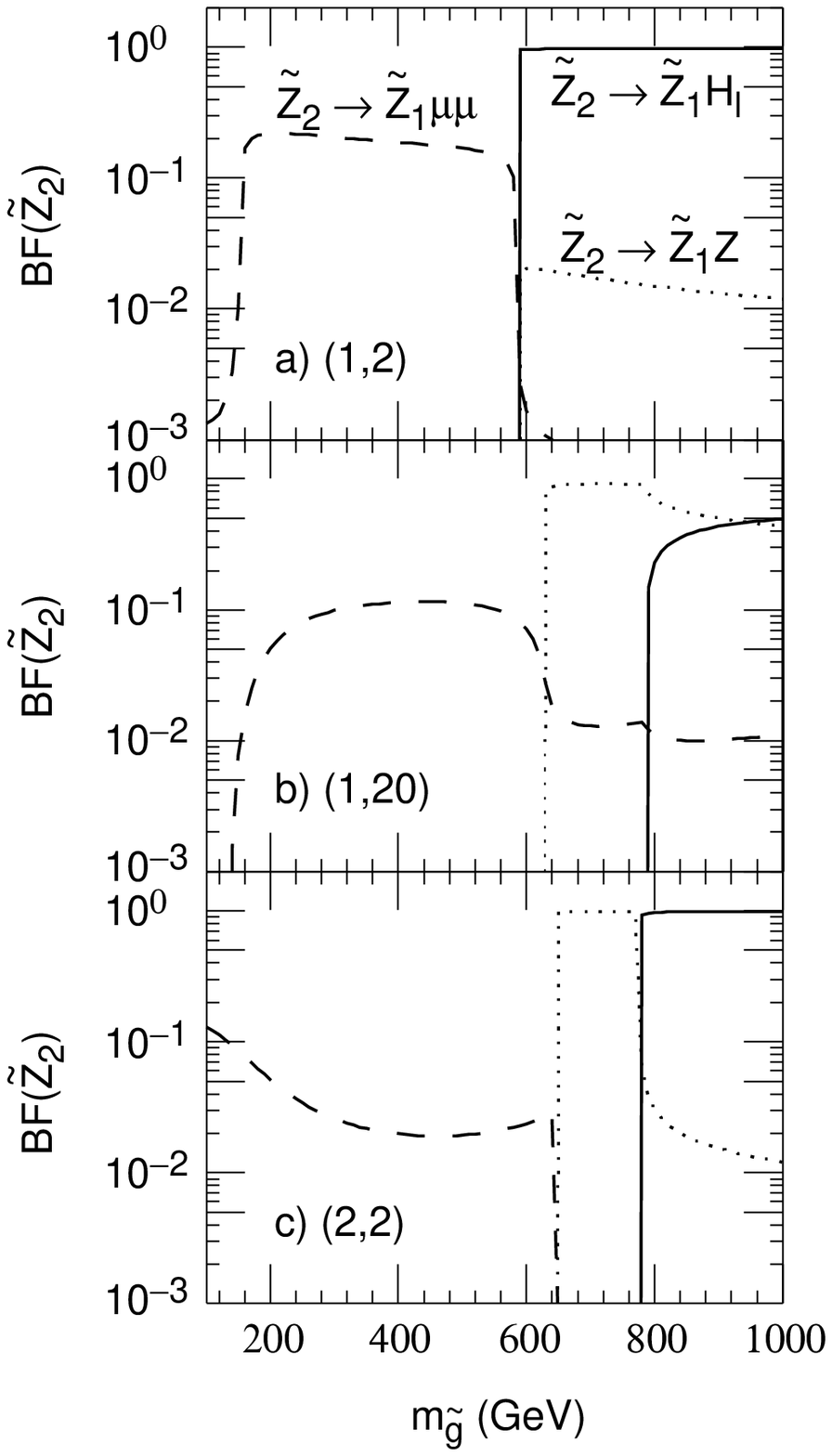}}
\bigskip\bigskip
\centerline{FIG.~3}
\vfill\eject

\centerline{\epsfbox{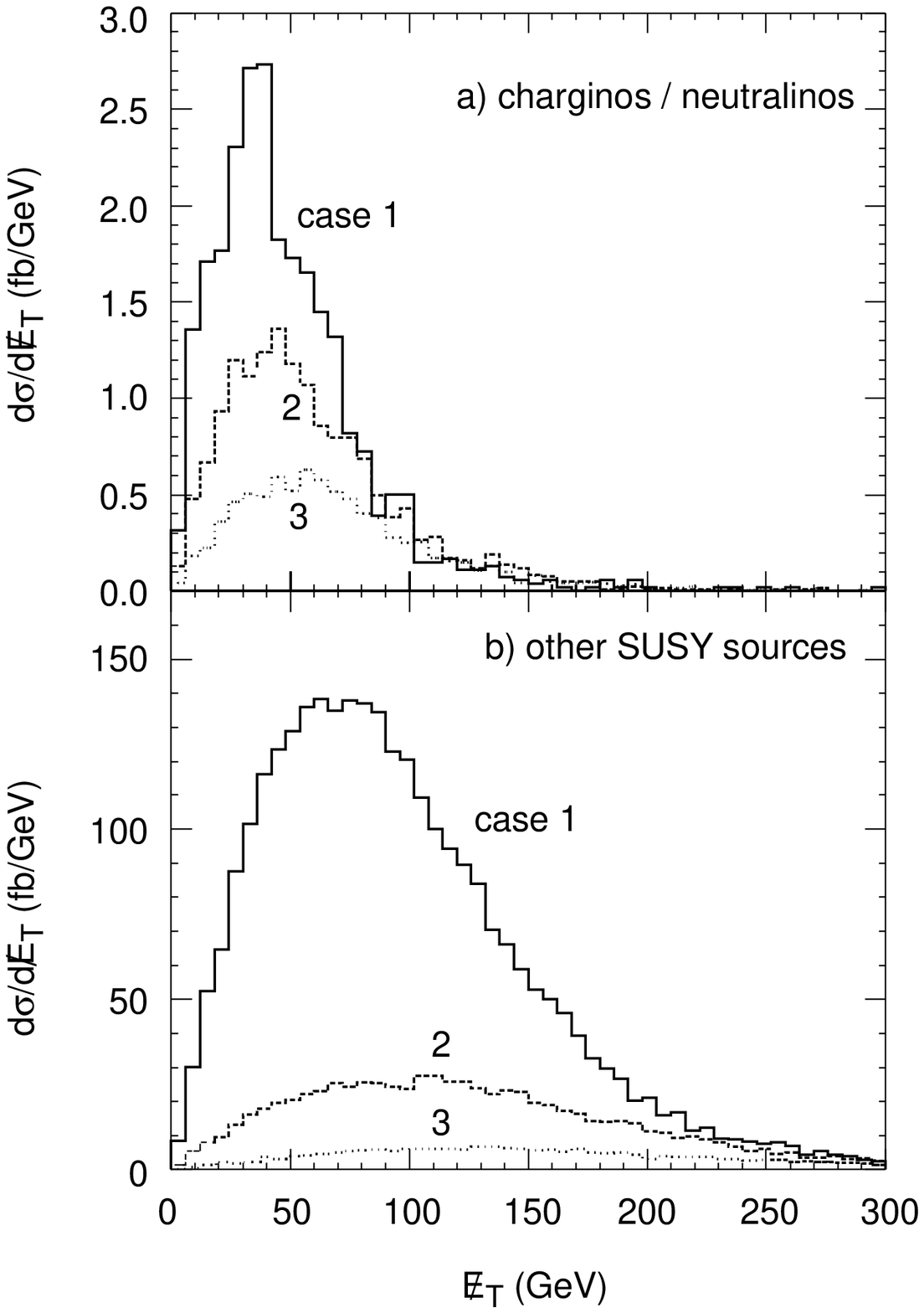}}
\bigskip\bigskip
\centerline{FIG.~4}
\vfill\eject

\epsfxsize=7in
\centerline{\epsfbox{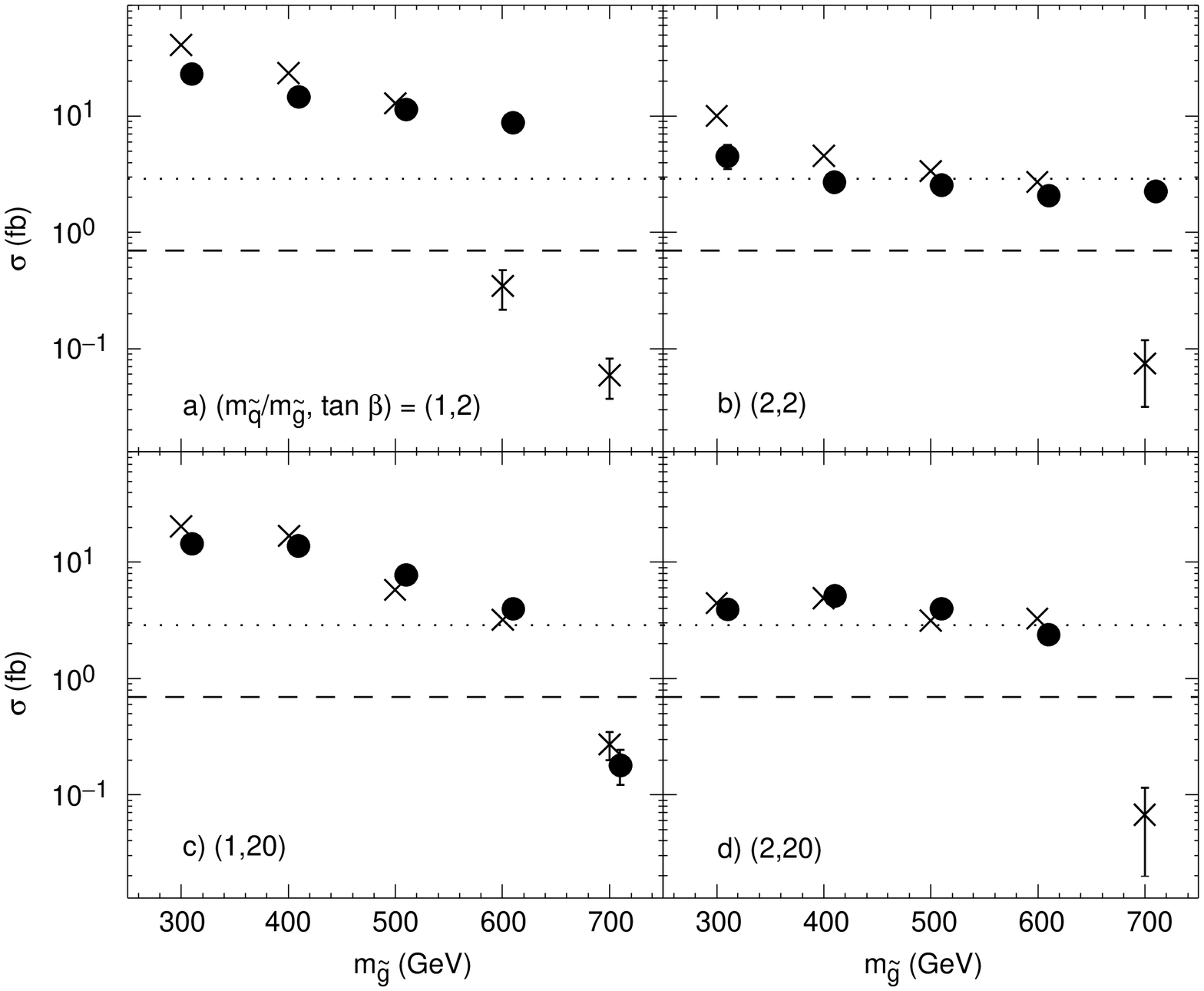}}
\bigskip\bigskip
\centerline{FIG.~5}
\vfill\eject

\epsfxsize=4.5in
\centerline{\epsfbox{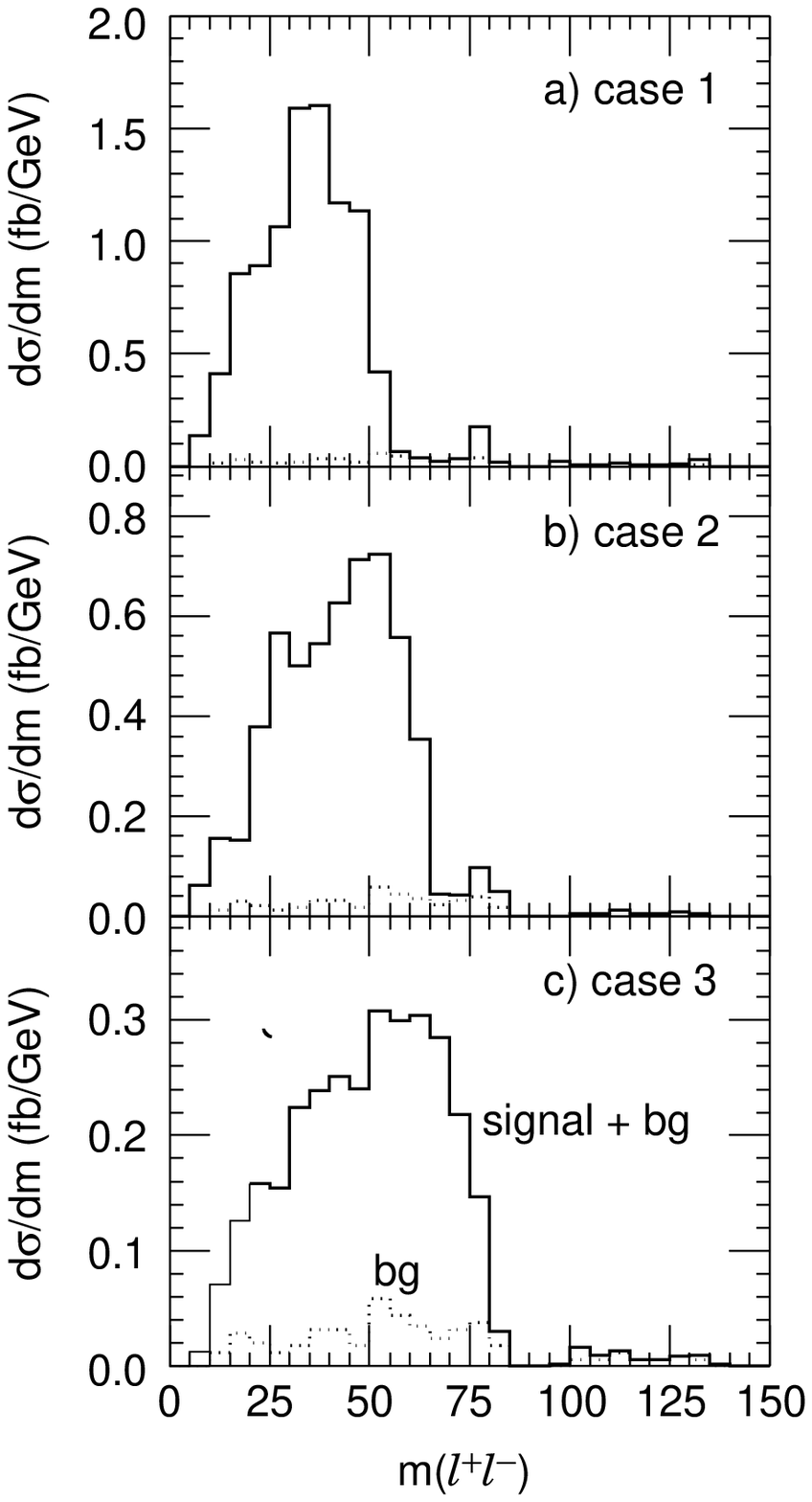}}
\bigskip\bigskip
\centerline{FIG.~6}
\vfill\eject

\centerline{\epsfbox{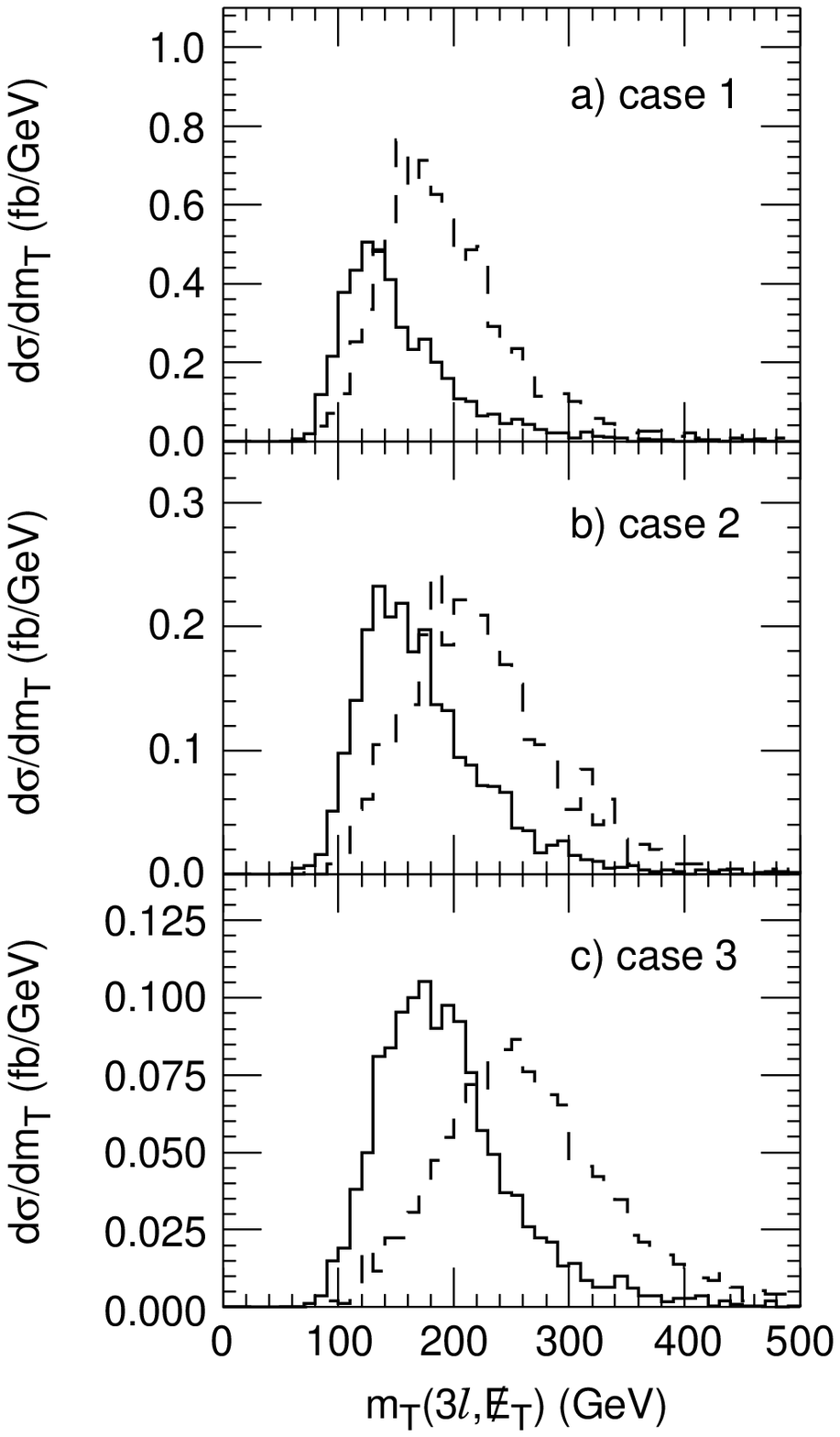}}
\bigskip\bigskip
\centerline{FIG.~7}
\vfill\eject

 \end{document}